# Decomposition of general grain boundaries

## Author information

Wei Wan [1], Junwen Deng [2] and Changxin Tang [1, *]

[1] *Institute of Photovoltaics, Nanchang University, Nanchang, 330031, China*

[2] *School of Advanced Copper Industry, Jiangxi University of Science and Technology, Yingtan 335000, China*

[*] *Corresponding author, Email address: tcx@ncu.edu.cn*

## Abstract

As a central part of microstructure evolution, grain boundary (GB) migration is believed to be both monolithic and unidirectional. But here, we introduce the concept of GB decomposition: one GB separates into two new GBs by exerting differential Peach-Koehler forces on its disconnections. Molecular dynamics simulation is used to reveal the disconnection mechanisms and direction-dependent motion behaviors associated with the reversible decomposition of a nickel Σ7 general GB. We also observed a decomposition-like process in a high-energy diffraction microscopy (HEDM) dataset of high purity nickel polycrystal (*Science* 2021, 374, 189–193), and performed HEDM-data-based simulations to confirm it. The decomposition should be considered as a new GB kinetic behavior, based on its particularity and potential universality.

## Introduction

Microstructure evolutions of crystalline materials are controlled by the kinetic process of defects, where the motion of grain boundaries (GBs) has received considerable attention due to its dominant role in plastic deformation, grain growth, recrystallization, first-order phase transformation, etc. [1–8]. Various mechanisms of GB migration, such as normal migration, shear-coupling, sliding and those that incorporate grain rotation, were investigated in numerous relevant studies [9–18]. These knowledges were soon applied in engineering and manufacturing to mediate the desired microstructural states and substantially increase the performance of crystalline materials [19, 20]. For example, the well-known Hall-Petch relationship [21–23] indicates that small grain sizes favor strength. Therefore, nanocrystalline metals with high GB content introduced will exhibit unique mechanical performances, but their microstructures are sometimes unstable because GBs will merge to reduce excess energy [24, 25]. Figures 1a and 1b show a sketch where a driving force $\boldsymbol{F}$ causes two GBs to migrate towards each other and merge. While this sketch is supported by many experimental observations, its inverse, i.e., a reversed driving force $-\boldsymbol{F}$ causes the merged GBs to separate, is never systematically confirmed.

Exploring the microscopic mechanisms of GB migration would solve this uncertainty. In Figure 1c, the disconnection (a defect with step height and dislocation characteristics) theory [26–31] elaborates the motion of a GB as: (1) disconnection nucleation across the GB and (2) disconnection glide along the GB. This theory also works when GBs are subjected to the constraints of the GB network and triple junctions in polycrystal [31]. However, both the disconnection theory and the kinetic equation of GB migration $\boldsymbol{v} = \boldsymbol{MF}$, (where $\boldsymbol{v}$ is GB velocity and $\boldsymbol{M}$ is mobility) [32, 33] do not prohibit such an assumption to reverse the GB merging (c.f. Figure 1d): through applying differential Peach-Koehler forces, $-\boldsymbol{F}$ will cause two disconnection types in a general GB to nucleate and glide oppositely, so that two separated GBs and a new grain between them are finally formed. We define this process as the decomposition of GBs.

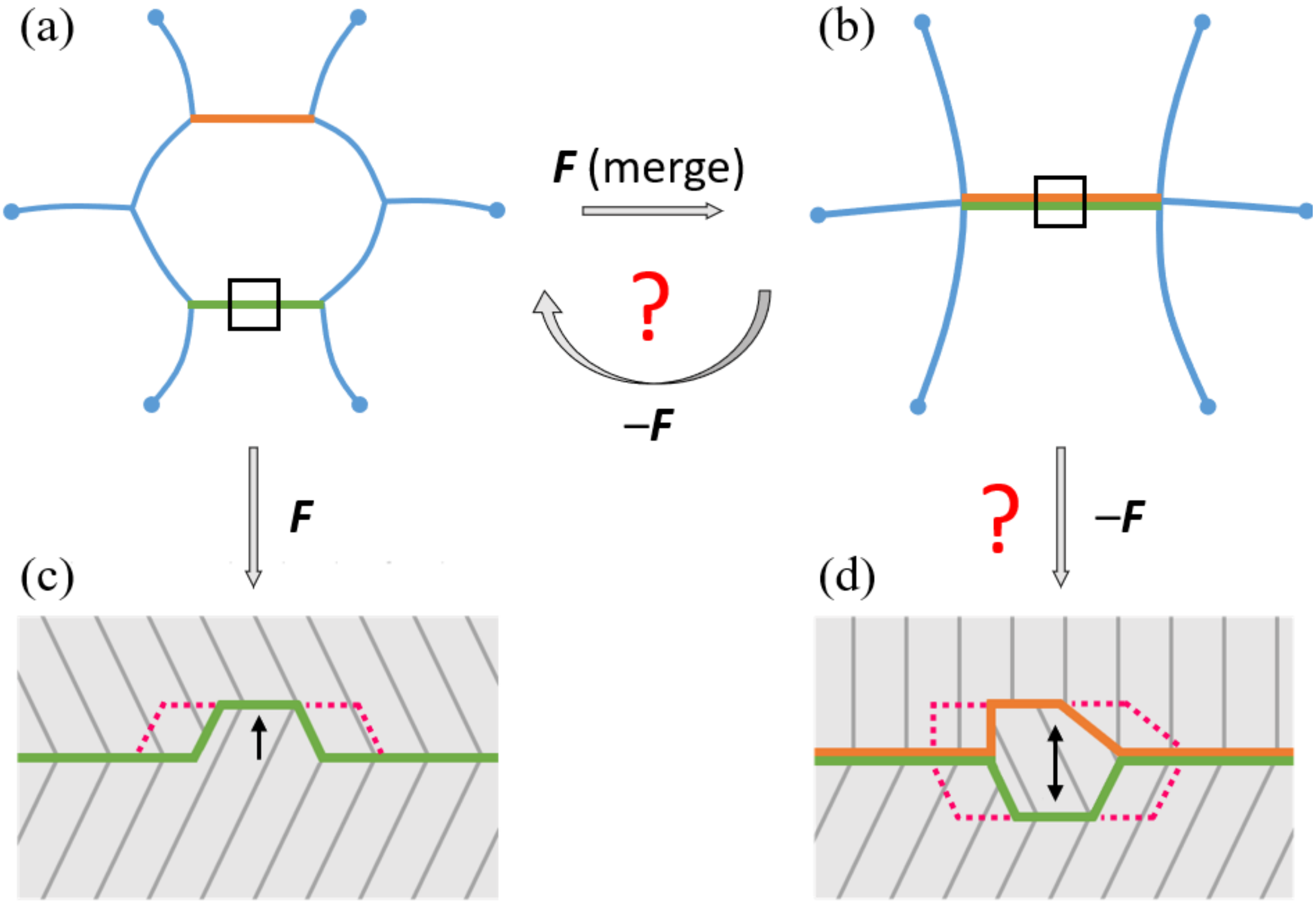

**Figure 1.** (a) GB network in polycrystals; (b) Merging of the orange and green GBs under the driving force; (c) GB migration in the disconnection theory; (d) An assumption showing the merged GB separates into the original GBs by simultaneous nucleation of two opposite disconnection types.

In this work, molecular dynamics (MD) simulation was adopted to explore the described decomposition phenomenon in a general GB dataset. The signatures of this phenomenon were analyzed in terms of structure, energy and stress. With the support of experimental data, the occurrence of decomposition under the constraint of the polycrystal environment is discussed.

## Decomposing a grain boundary

The fundamental principle of decomposing a given GB is, to elaborate a type of driving force or stress state (including chemical potential jump driving force, directional tensile/compressive forces, directional shear forces and/or their combinations) that exerts different and desired Peach-Koehler forces [34, 35] on different dislocations or disconnections, as their resolved shear stresses. A prototype example of how to decompose and decomposition of the low angle GBs was given in our previous work [36]. Such experiences are extended to high angle general (mixed tilt-twist) GBs in this work, because the nature of decomposition requires at least two disconnection types and thus forbids the commonly seen simple GBs (e.g., symmetric tilt GB) from decomposition. The overall procedures are described in the *Supplementary Materials* and briefly stated below.

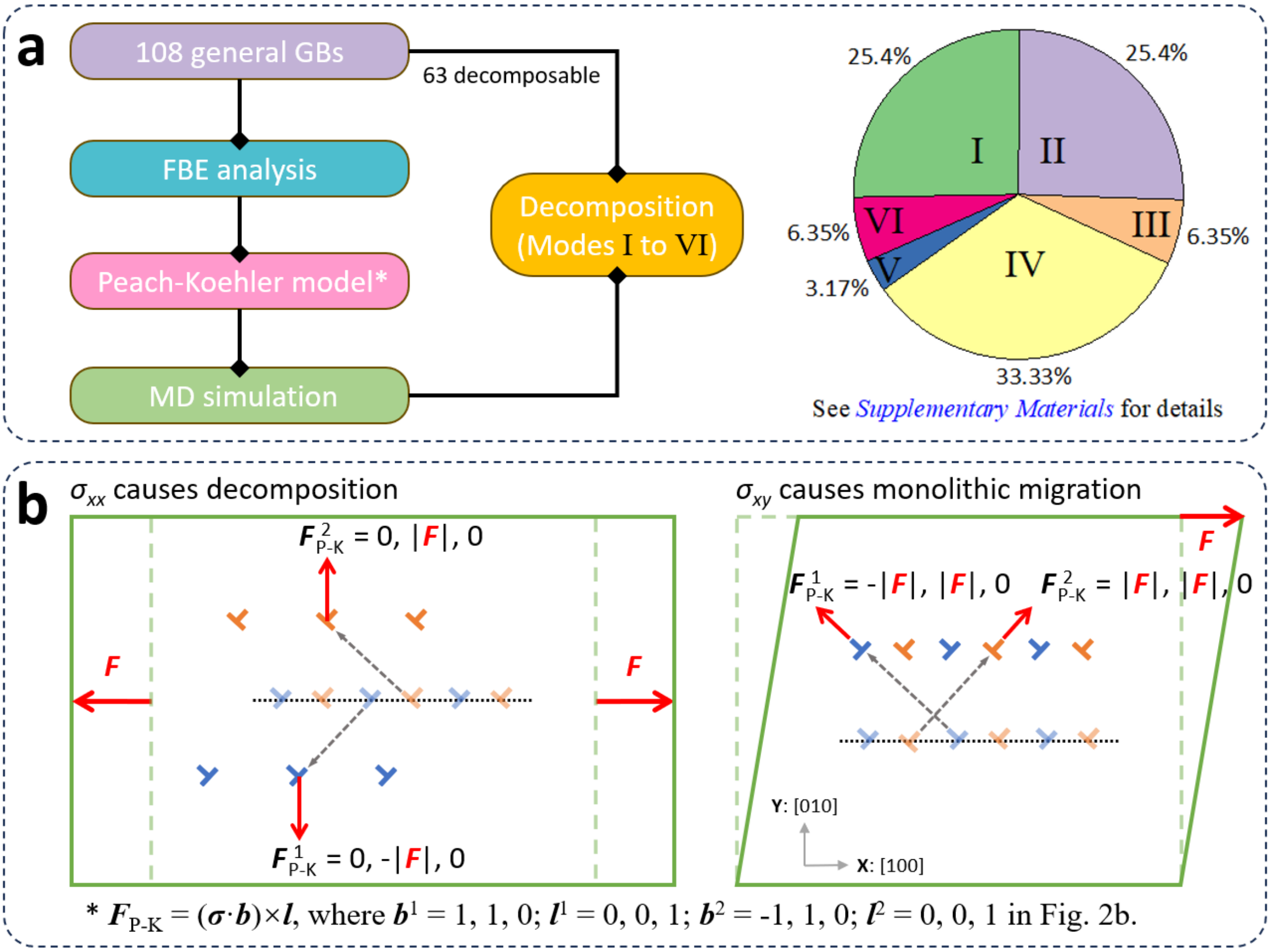

**Figure 2.** (a) Working flow to investigate the decomposition of GBs, where six decomposition modes are observed in 63 decomposable GBs; (b) A GB with two dislocation/disconnection types will respond differently to directional tensile and shear forces due to the exerted Peach-Koehler forces on its dislocations/disconnections.

The potential GB decomposition process in FCC nickel was investigated via MD simulation, because the relevant computational and experimental datasets are abundant [37–43] for reference. Based on the previous studies [40–43] about nickel GB migration, the Foiles-Hoyt interatomic potential [44] was employed to model the interaction of nickel atoms. The working flow of the investigation is presented in Figure 2a. First, to prove the decomposition is not limited to isolated cases, we picked 108 general GB characters and sample the corresponding GB structures with minimum excess energy through a well-established sample method [45–48] implemented with the LAMMPS software [49]; Second, we solved the disconnection characteristics of the GBs from their macroscopic characters, which involves the use of the Frank-Bilby equation (FBE) [50–53] and analysis of the GB structure [54]. Third, we applied the Peach-Koehler model to determine the type of driving force that may cause the GB to decompose, an example of which is plotted in Figure 2b. Then, a constant driving force of that type was exerted on the relaxed GB structure under 300 K for sufficient times until the decomposition occurred. Fourth, the simulation reported that 63 GBs can decompose and 41 of them have low $\Sigma$ value ($\Sigma < 100$), indicating the universality of the decomposition across various GB characters and structures. Notably, six decomposition modes incorporating different structural mechanisms were observed, see Figure 2a and *Supplementary Materials*. However, we chose to analyze one representative case to illustrate the concept of GB decomposition, and leave these for an in-depth study in the future.

## Results

### Simulated decomposition in bicrystal

With the aid of the analysis tool OVITO [55], the decomposition of a Σ7 general GB is selected to present, due to its relatively simple disconnection mechanisms. The atomic structure of this GB is given in Figure 3a, where the disconnections are marked and the driving force type for decomposition is the directional shear force $\tau_{xz}$. After applying a constant driving force magnitude $|\boldsymbol{F}| = |\tau_{xz}| = 1.48$ MPa for 144 ps, Figures 3b and 3c capture two disconnection types nucleating and gliding on both sides of the GB plane. In Figure 3d, the two disconnection types fully separate and hence form two new GBs, which migrate oppositely to reorient the atoms between them. A new grain thus emerges and grows in Figure 3e. This decomposition also shares a signature with the detwinning [56], i.e., creating stacking faults that are identified as HCP atoms. Nudge elastic band (NEB) calculations [57, 58] in Figures 3f and 3g reveal the energy barriers during the process: 66 mJ/m$^2$ is the barrier to activate simultaneous nucleation and glide for decomposition, while 21 mJ/m$^2$ must be overcome to grow the new grain. Comparing with the existing barriers of GB migration [13], our calculation results are rational and suggest that the decomposition is a rare event.

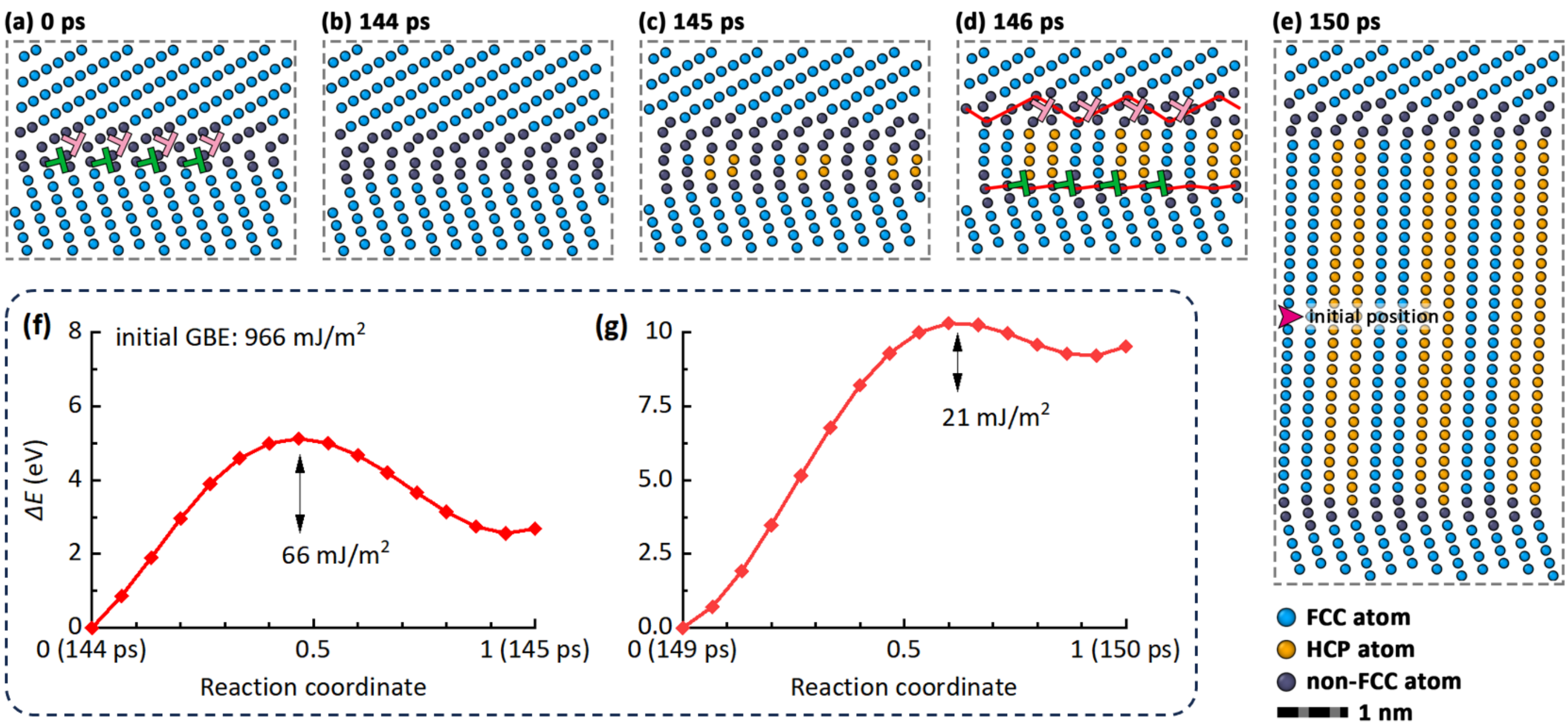

**Figure 3.** MD simulated decomposition process of a nickel Σ7 general GB under 300 K. (a), (b), (c), (d) and (e): Snapshots of the GB structure capturing the decomposing process at different times; Disconnections are marked by a colored symbol ⊥ to show their motions; Conjugate-gradient energy minimization is used to eliminate the thermal noises; (f) and (g): NEB calculated minimum energy paths, where (f) is from the structures before (144 ps) to after the decomposition (145 ps), and (g) is from the structures before (149 ps) to after the growth of the new grain (150 ps). Energy barriers are converted to the unit of energy per area.

To further examine the stability of the new grain and the reversibility of decomposition process, we cancel $\boldsymbol{F}$ after the two GBs reached their steady velocities for a while, and reverse the direction of $\boldsymbol{F}$. For convenience, the migration and geometry parameters of GBs and grains are defined in Figure 4a. The vertical displacements of GBs #1 and #2 ($D_{\mathrm{V}}^{\#1}$ and $D_{\mathrm{V}}^{\#2}$) versus time are recorded in Figure 4b, where the GB vertical velocities are calculated by fitting the curve. The first valley of $D_{\mathrm{V}}^{\#1}$ corresponds to a rotation of the newly emerged grain N at 155 ps (see *Supplementary Materials* for details); both GB velocities are steady then, i.e., -116 m/s and 45 m/s. At 300 ps, $\boldsymbol{F}$ is cancelled, lasting 200 ps, so that the two GBs would stop. System stability is also confirmed since the size of grain N does not shrink in this period. Meanwhile, the calculation of grain orientations, presented in Figure 4a, indicates that GBs #1 and #2 are metastable GBs with strain/long-range stress. Reversing $\boldsymbol{F}$ to $-\boldsymbol{F}$ at 500 ps makes two GBs migrate reversely, as the classical equation $\boldsymbol{v} = \boldsymbol{MF}$ regulates. Note that both new GBs exhibit direction-dependent motion behaviors, e.g., the absolute GB velocities driven by the opposite driving forces are not the same, which results in a nonzero net displacement for the original GB #0 after GBs #1 and #2 merged.

Figure 4c plots the lateral displacements of grains 1 relative to 2, grains 1 relative to N and grains N relative to 2 ($D_{\mathrm{L}}^{1-2}$, $D_{\mathrm{L}}^{1-\mathrm{N}}$ and $D_{\mathrm{L}}^{\mathrm{N}-2}$, respectively), from which the reciprocal shear-coupled factors of GBs #1, #2 and #0 ($\beta_{\#1}^{-1}$, $\beta_{\#2}^{-1}$ and $\beta_{\#0}^{-1}$, defined as the GB vertical velocity divides the lateral velocity) are calculated. The results vary smoothly in the period when the two GB velocities are steady. $\beta_{\#0}^{-1}$ is zero after the two GBs merged, implying not only the occurrence of GB sliding, but also the direction-dependence of GB decomposition itself. As for the stress-time curve of Figure 4d, decomposition or sliding of GB #0 requires high shear stress near 4 GPa, but the stress to grow or shrink the size of grain N is much lower, which is consistent with the calculated energy barrier results.

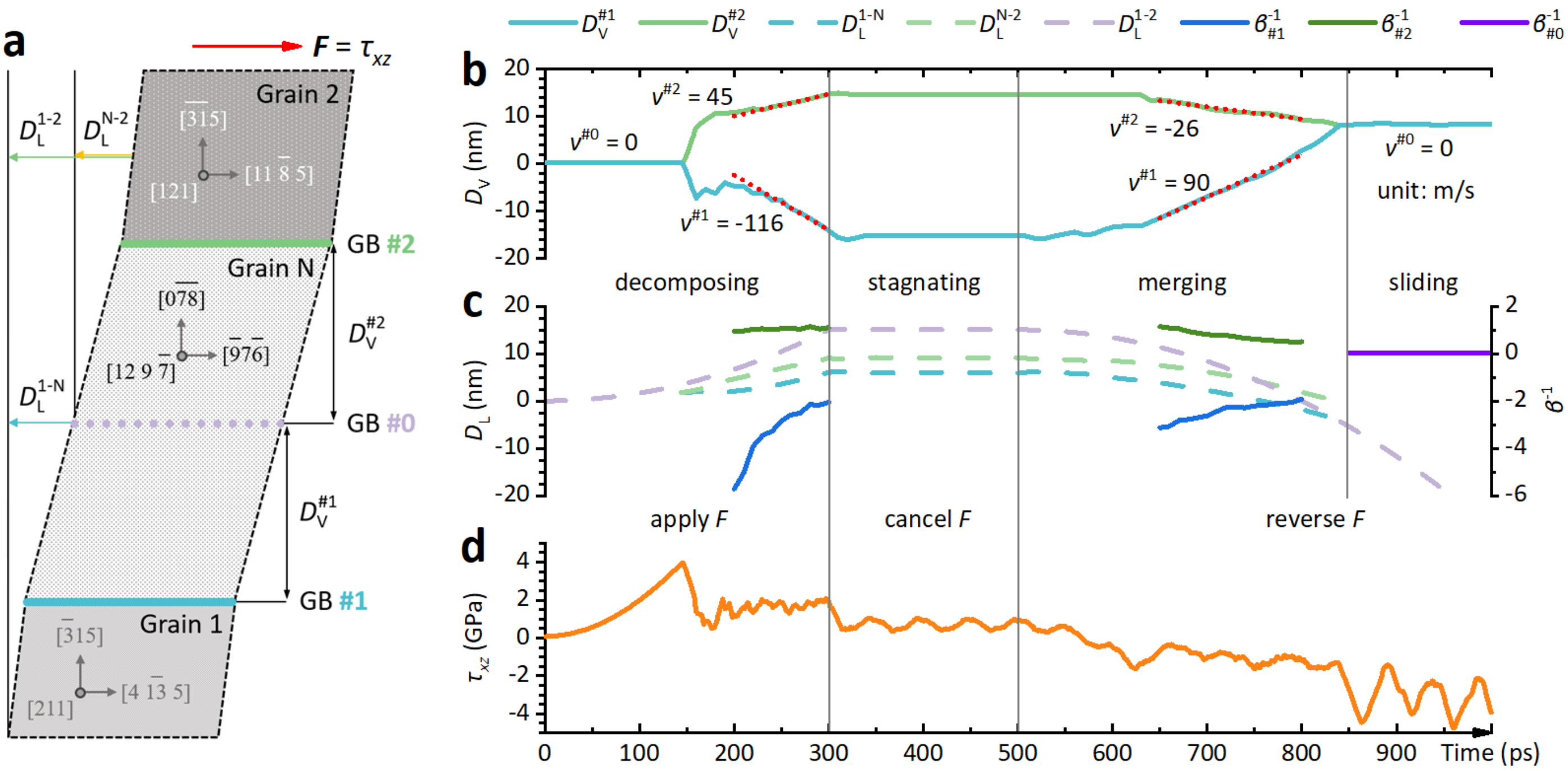

**Figure 4.** (a) Parameter definition and geometry setting of the GB decomposition; (b) Vertical displacements of GBs #1 and #2 versus time; (c) Lateral displacements between the two/three grains and reciprocal shear-coupled factors of GBs #0, #1 and #2 versus time; (d) Stress-time curve, $\tau_{xz}$ is the shear stress of the system.

### Experimentally observed decomposition in polycrystal

Reproducing this theoretically viable process in in-situ experiments would provide powerful evidence to support the simulation results. Due to the difficulties when performing a costly bicrystal-based experiment, we try to resort to an existing HEDM dataset of a high purity nickel polycrystal from Aditi Bhattacharya et al. [37], where the grain evolution during annealing under 1073K was recorded at different times. The pros of using this dataset is: (1) most high angle GB characters in polycrystal are already the general type satisfying the requirement of at least two disconnection types; (2) Our claims could be generalized if a decomposition occurs in polycrystal, because the constraint of the GB network may prevent the migration behaviors in bicrystal, such as the GB velocity-curvature relationship, to appear in polycrystal [37]; while the cons is: (3) the HEDM method only collects the information at μm-scale, and therefore difficult to explore the mechanisms at the atomic level.

By inspecting the inverse pole figures (IPFs), a new grain emerged at a GB after ≈30 minutes annealing is found in a polycrystal region (z = 25). The surrounding regions (z = 22 to 28) are also inspected to ensure this new grain does not originate from the growth of an existing grain nearby, see Figure 5. But what happened here should still be classified as a suspicious decomposition process, because the new grain may exist long before the annealing process due to the limitation of the HEDM resolution (2.3 × 2.3 × 4 $\mu m^3$) [37]. This is why we performed an HEDM-data-based MD simulation for mutual corroboration. Since the polycrystal system is μm-scale and MD simulation is limited to nm-scale, we intend to simulate a small bicrystal region with periodic boundary conditions, and use this local structure to represent a part of the polycrystal system.

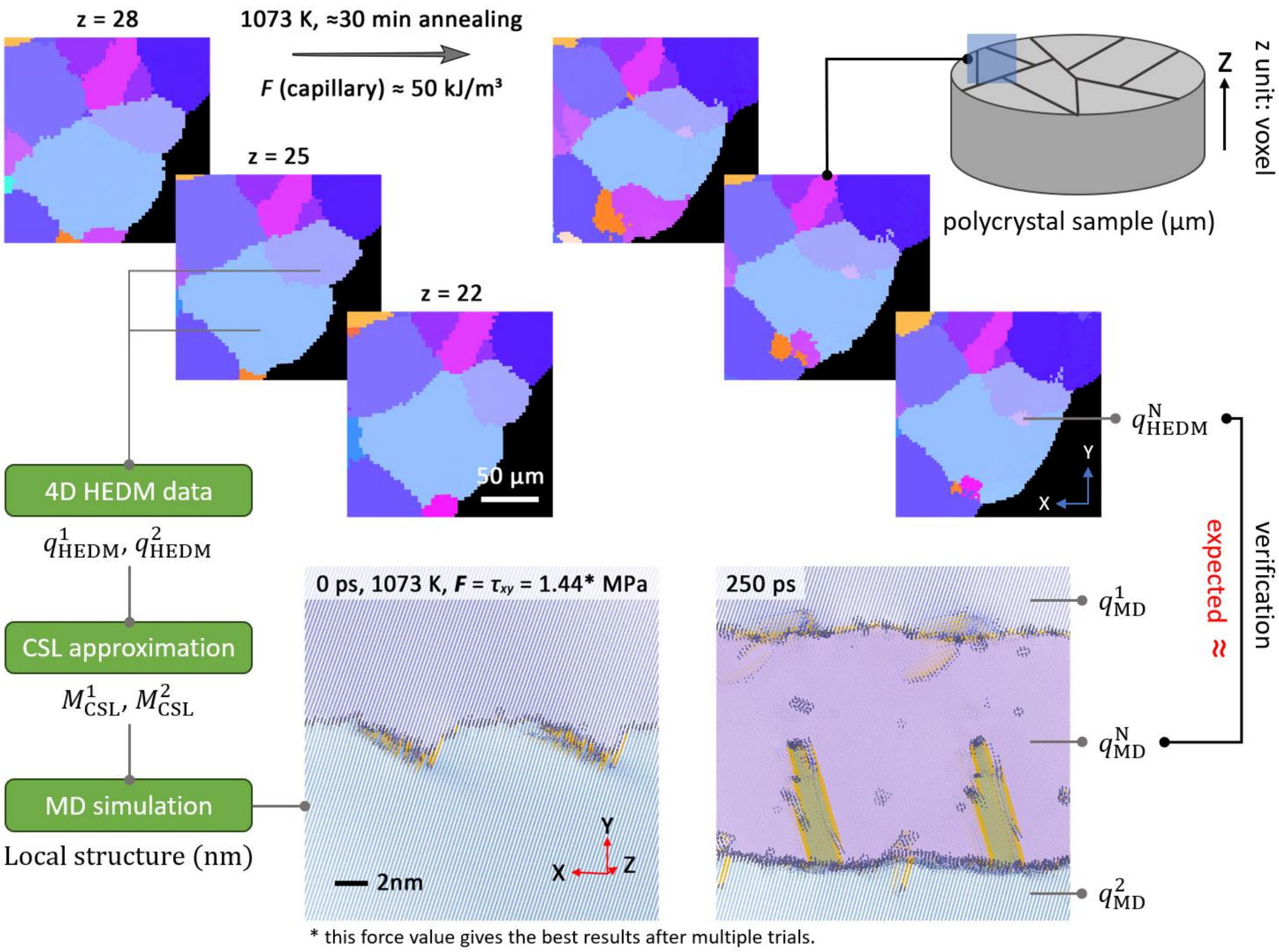

**Figure 5.** HEDM IPF records and MD simulation of a decomposition-like process in a high purity nickel polycrystal, and the working flow to simulate a GB from the HEDM data. FCC nickel atoms are colored by the grain quaternions to be consistent

with the IPFs. Color styles of the non-FCC atoms follow the same in Figure 2.

**Table 1.** Comparison and error measurement of quaternions obtained from the HEDM experiment [37] and MD simulation.

| Quaternion ID | Quaternion value [*x*, *y*, *z*, *w*] | Angular error (°) | Euclidean distance |
|---|---|---|---|
| $q^{1}_{\mathrm{HEDM}}$ | [−0.1353, −0.2993, 0.7805, 0.5319] | 7.29* | 0.0608 |
| $q^{1}_{\mathrm{MD}}$ | [−0.0993, −0.2616, 0.8094, 0.5163] | | |
| $q^{2}_{\mathrm{HEDM}}$ | [0.1753, −0.2917, 0.4749, 0.8115] | 3.62* | 0.0321 |
| $q^{2}_{\mathrm{MD}}$ | [0.1892, −0.3027, 0.4941, 0.7928] | | |
| $q^{\mathrm{N}}_{\mathrm{HEDM}}$ | **[−0.0093, 0.2922, 0.8147, 0.5007]** | **4.58** | **0.0742** |
| $q^{\mathrm{N}}_{\mathrm{MD}}$ | **[−0.0086, 0.3513, 0.8215, 0.4556]** | | |

* The error is largely introduced by the CSL (coincident site lattice) approximation, where we try to select low Σ GB characters for approximation as possible due to their frequent occurrences in polycrystal, see *Supplementary Materials* for details.

The logic of mutual corroboration between simulation and experiment is summarized in Figure 5: First, from the Dream.3D software [59], we get quaternions $q^{1}_{\mathrm{HEDM}}$ and $q^{2}_{\mathrm{HEDM}}$ that specify the grain orientations in the region, from which accurate experimental GB character is obtained; Second, as the GB character is non-CSL type, its rotation matrices will be approximated by integer transformation matrices $M^{1}_{\mathrm{CSL}}$ and $M^{2}_{\mathrm{CSL}}$ to obtain a crystallographic-close low Σ CSL GB character. Third, since the actual driving force on that region could not be interpreted from the HEDM dataset, the working flow in Figure 2 is reused to try to decompose this CSL GB. If it decomposes and the simulation reproduces the HEDM observation, then the quaternion $q^{\mathrm{N}}_{\mathrm{MD}}$ of the new grain N obtained from the MD simulation should be very close to its experimental counterpart $q^{\mathrm{N}}_{\mathrm{HEDM}}$. In the *Supplementary Materials*, more details and analysis about this HEDM-data-based simulation are given.

MD simulation results under the experimental annealing condition are also presented in Figure 5. The CSL GB used for approximation has a facet structure, and decomposes under constant shear force $\tau_{xy}$. Here, we focus on verifying the resultant new grain that emerged from the decomposition, because the huge difference in the applied strain rate could not guarantee the reproduction of the experimental driving force and GB velocity in the MD simulation. Quaternions of each grain in the MD simulation ($q^{1}_{\mathrm{MD}}$, $q^{2}_{\mathrm{MD}}$ and $q^{\mathrm{N}}_{\mathrm{MD}}$) are computed and compared with the HEDM data in Table 1. Notably, due to the low Euclidean distances, these quaternions are nearly consistent, which indicates that the emergence of new grain in that polycrystal region is indeed a GB decomposition process. On the other hand, the evolution of nickel polycrystal, recorded by these valuable HEDM IPFs, suggests that the real general GB could decompose too. While GB merging processes are frequently observed in this polycrystal, we only captured very few cases of decomposition; such scarcity may come from (1) the constraint of the GB network and (2) the high energy barrier to activate the decomposition.

## Conclusions & Discussions

This work shows that a general GB could be decomposed into two new GBs by controlling the Peach-Koehler forces on its disconnections. Both simulated and experimentally observed decomposition processes are investigated. The conclusions are summarized below:

(1) Simulating the decomposition of a nickel Σ7 GB reveals its atomic scale mechanism as different disconnection nucleation on both GB sides, which overcomes a high energy barrier to form two independent metastable GBs with a new reoriented grain emerging. Examination of the reversibility of this process unveils direction-dependent behaviors during the growth and shrinking of the new grain. The Σ7 GB itself is also a Brownian ratchet [13] because it responds to a given shear force with decomposition and turns to the sliding under the reversed shear force.

(2) In a HEDM dataset about nickel polycrystal, a similar process is observed, where a new grain emerges from an existing GB. The local GB structure is approximately reproduced in the MD simulation, which subsequently confirms the process as a GB decomposition. Mutual corroboration between simulation and experiment is achieved by comparing the obtained grain quaternions, and their consistency not only suggests our simulation is reliable, but also indicates the viability of the GB decomposition even under the constraint of a polycrystal GB network.

We are not the first to show one GB becomes to two, such as the dislocation dissociation of low angle GBs [60, 61] and triple junction reconstruction [62], but we attempt to propose and access the concept of decomposition through a joint application of the FBE, the Peach-Koehler model and MD simulation, which is a universal methodology viable for any given GB. Given the fact that this concept is verified in polycrystals, many theories about GB migration are being challenged. The first should be the classical equation $\boldsymbol{v} = \boldsymbol{MF}$, where the mobility $\boldsymbol{M}$ is often considered as a tensor [63, 64]. However, if we redefine $\boldsymbol{M}$ for disconnections as their properties, we can easily explain the GB decomposition while considering all conventional GB migrations as the interactions between independent disconnection migrations. For example, assuming a GB has two disconnection types with driving-force-dependent 1D velocity $v_1$ and $v_2$, if the signs of $v_1$ and $v_2$ are opposite, then the GB will decompose. Otherwise, if $v_1 > v_2 > 0$, then both disconnections should move together in an overall velocity $v$ where $v_2 < v < v_1$ (expected). This viewpoint is also discussed in our previous work [36], where the dragging effects between the mobile and immobile GBs are more obvious during the decomposition of low angle GBs.

We should also reevaluate the simulation methodology of GB migration. This is because we are not clear whether a chemical potential jump would activate the GB decomposition in real materials. Contradictorily, any potential GB decomposition triggered by the chemical potential jump will not be captured by the current synthetic driving force-based MD simulation in principle due to its bicrystal-based implementation [65, 66].

Constraints of the GB network should not affect the occurrence of decomposition in a designated polycrystal structure, according to the Peach-Koehler model theoretically. In contrast to the frequently observed GB merging process, addressing the reason that causes the scarcity of the decomposition is important. Beyond the Peach-Koehler model, the local stress driving the disconnections to separate is expected to be high enough to overcome the barrier. This naturally requires strong directional force conditions on local structures, which may not be satisfied in most polycrystal experiments. Furthermore, even if a directional force is applied to the polycrystal, it will first drive the GBs with low barriers to migrate. The stress is dissipated to these easily mobile GBs instead of activating decomposition, so that the decomposition is a rare event and experimentalists barely notice its existence.

## Data availability

Numerical data are available upon reasonable request.

## Acknowledgements

W. Wan acknowledges the fruitful discussions with Prof. E.R. Homer, who has helped to establish the concept of grain boundary decomposition and the application of the Peach-Koehler model. W. Wan also acknowledges Prof. J.B. Yang for the use of the Frank-Bilby equation and the discussion about dislocation dissociation in low angle grain boundaries. W. Wan and C.X. Tang acknowledge the computational resource support from Institute of Aerospace Research at Nanchang University.

## Competing interests

The authors declare no competing interests.

## Fundings

(1) W. Wan acknowledges the financial support from the Department of Mechanical Engineering, Brigham Young University.

(2) C.X. Tang was supported by the National Natural Science Foundation of China (grant number: 12464044).

## Author contributions

(i) W. Wan carried out this project, performed all simulations, visualized all figures, established the procedure to decompose a grain boundary, wrote the original manuscript and revised it.

(ii) J.W. Deng helped to find the decomposition-like process in the HEDM IPF dataset, and partially engaged in the writing of *Supplementary Materials*.

(iii) C.X. Tang supervised the project, contributed computation resources, maintained the data and revised the manuscript.